\documentclass[12pt,english,aps]{revtex4}

\usepackage[T1]{fontenc}
\usepackage[latin9]{inputenc}
\usepackage[a4paper]{geometry}
\usepackage{amsmath}
\usepackage{graphicx}
\usepackage{amssymb}
\usepackage{esint}
\usepackage{babel}

\begin{document}

\title{Observation of Grand-canonical Number Statistics in a Photon Bose-Einstein condensate}

\author{Julian Schmitt, Tobias Damm, David Dung, Frank Vewinger, Jan Klaers and Martin Weitz}

\address{Institut f\"ur Angewandte Physik, Universit\"at Bonn, Wegelerstr.
8, 53115 Bonn, Germany}

\begin{abstract}
We report measurements of particle number correlations and fluctuations of a photon Bose-Einstein condensate in a dye microcavity using a Hanbury Brown--Twiss experiment. The photon gas is coupled to a reservoir of molecular excitations, which serve as both heat bath and particle reservoir to realize grand-canonical conditions. For large reservoirs, we observe strong number fluctuations of the order of the total particle number extending deep into the condensed phase. Our results demonstrate that Bose-Einstein condensation under grand-canonical ensemble conditions does not imply second-order coherence.
\end{abstract}
\maketitle
Large statistical number fluctuations are a fundamental property known from the thermal behavior of bosons, as has been strikingly revealed in Hanbury Brown--Twiss experiments with both photons and material particles \cite{hanbury1956,baym1998,Schellekens2005,Jeltes2007,Oettl2005,Hodgman2011,Perrin2012,Dall2013,Deng2010}. At low temperatures and high densities however, when a macroscopic fraction of bosonic particles undergoes Bose-Einstein condensation (BEC) 
\cite{Einstein1925}, large fluctuations of the condensate population conflict with particle number conservation \cite{Kocharovsky2006}. Correspondingly, condensation is in this case accompanied by a damping of fluctuations and the emergence of second-order coherence \cite{Schellekens2005,Jeltes2007,Oettl2005}. Experimentally, Bose-Einstein condensation has been achieved with cold atomic gases and solid-state quasiparticles, exciton-polaritons and magnons, respectively \cite{Anderson1995,Davis1995,Deng2010,Demokritov2006,Bennemann2013}. Moreover, in recent work we observed the Bose-Einstein condensation of a two-dimensional photon gas in a dye microcavity \cite{Klaers2010}, where photons thermalize with dye molecules by repeated absorption-emission processes. Because of the frequent interconversion of photons with molecular electronic excitations, the dye in this system acts as both heat bath and particle reservoir. The latter allows for fluctuations of the particle number around a (controllable) average value, a situation that for large reservoirs is well described in terms of the grand-canonical statistical ensemble \cite{Klaers2012}.

Here, we report measurements of correlations and counting statistics of a photon Bose-Einstein condensate coupled to a particle reservoir of variable size. We observe a regime with condensate number fluctuations of order of the total particle number, which demonstrates Bose-Einstein condensation under grand-canonical statistical conditions. Such a situation has not been studied previously in any other quantum gas experiment; e.g., typical ultracold atomic gas experiments realize microcanonical conditions. Furthermore, when decreasing the reservoir size, a crossover to a regime with reduced, Poissonian-type number fluctuations is observed.

In statistical physics, different statistical ensembles represent different laws of conservation that can be realized in nature. The microcanonical ensemble describes a system with energy and particle number strictly fixed, while in the canonical ensemble energy is allowed to fluctuate around a mean value determined by the temperature of a heat reservoir. Under grand-canonical conditions, both an exchange of energy and particles with a much larger reservoir is allowed, leading to fluctuations in both quantities. For large systems, the different statistical approaches usually become interchangeable \cite{Huang1987}, in the sense that relative fluctuations should vanish in all of them, i.e., $\delta N/N \rightarrow 0$ for the average total particle number $N$ and its root mean square deviation $\delta N$. This assumption is violated in the grand-canonical treatment of the ideal Bose gas, where the occupation of any single particle state suffers relative fluctuations of $100\%$. Applied to the macroscopically occupied ground state of a Bose-Einstein condensed gas, this implies statistical fluctuations of the order of the total particle number \cite{Kocharovsky2006,Fierz1955,Fujiwara1970,Ziff1977,Wilkens1997,TerHaar1977,Navez1997,Holthaus1998}, i.e., $\delta N \simeq N$. Whereas one usually expects fluctuations to freeze out at low temperatures, here the reverse situation is encountered: the total particle number will start to strongly fluctuate as the condensate fraction approaches unity, a behavior termed grand-canonical fluctuation catastrophe \cite{Kocharovsky2006,Holthaus1998}.

However, the physical significance of the grand-canonical ensemble for experiments below the condensation temperature has long been disputed \cite{Ziff1977,Wilkens1997,TerHaar1977}. Ziff, Uhlenbeck and Kac showed that the grand-canonical ensemble loses its validity for a system in diffusive contact with a spatially separated particle reservoir \cite{Ziff1977}. Their arguments, though, do not apply for system and reservoir that are spatially overlapped and consist of different species (or internal states) subject to particle (or states) interconversion. This situation is realized in the case of the BEC of a photon gas in a dye-filled microcavity, where photons are frequently absorbed and emitted by dye molecules \cite{Klaers2010,Klaers2012,Klaers2010th}. This process can be seen as a photochemical reaction $\gamma\ + \downarrow\ \rightleftarrows\ \uparrow$, where photons $(\gamma)$, ground state $(\downarrow)$, and excited dye molecules $(\uparrow)$ are repeatedly converted into each other, and the molecules work as a Òreservoir speciesÓ. Here, it is not the number of particles (photons) that is conserved but rather the total number of photons and molecular excitations. This allows us to experimentally explore grand-canonical statistics in the condensed phase.

\begin{figure}
\includegraphics{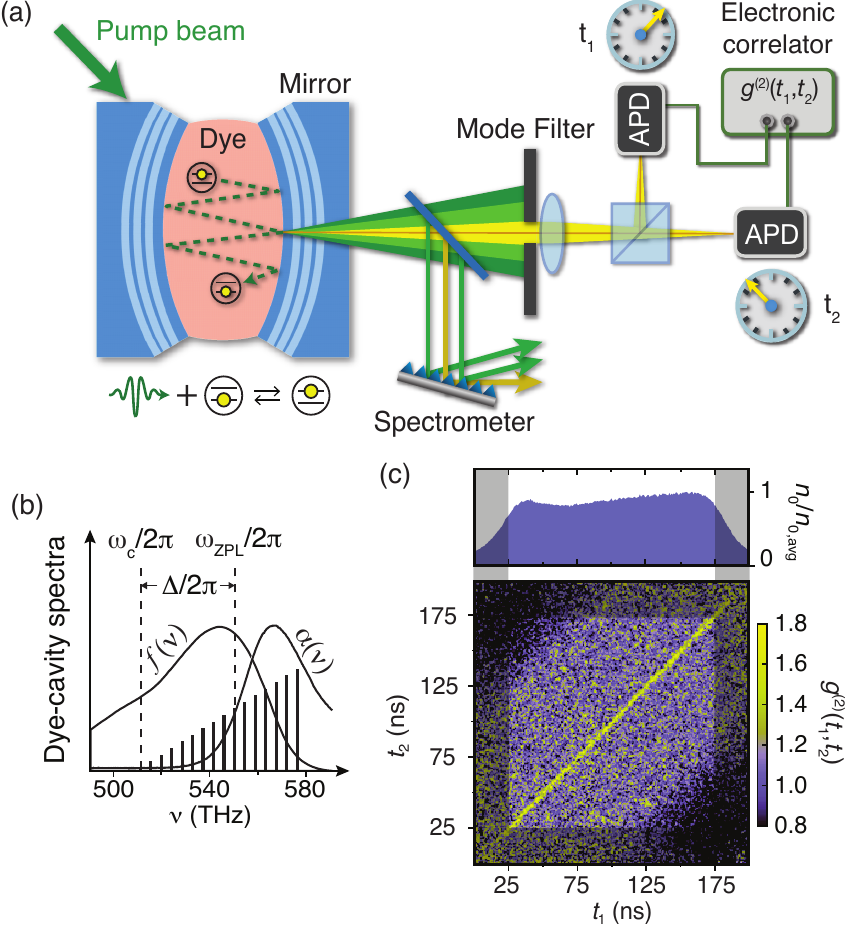}
\caption{(a) Experimental setup. Inside the dye-filled microresonator, photons couple to the dye reservoir by repeated absorption and emission, corresponding to a photochemical reaction as indicated below the resonator. Condensate light is separated by mode filtering, and directed onto a Hanbury Brown--Twiss set-up to measure the photon statistics. (b) Relevant transverse cavity modes TEM$_{8mn}$ (height of bars indicating degeneracy, cavity cutoff $\omega_\textrm{\tiny c}$) superimposed onto the rhodamine 6G absorption $\alpha(\nu)$ and emission $f(\nu)$ spectral profile (dye zero-phonon line $\omega_\textrm{\tiny ZPL}$). (c) Typical experimental results for the average condensate population during one measurement cycle (top) and the second-order correlation function $g^{(2)}(t_1,t_2)$ (bottom).
\label{fig1}}
\end{figure}

Our experimental setup [Fig.~\ref{fig1}(a)] is similar to that described in previous work on the realization of a photon Bose-Einstein condensate; see Refs.  \cite{Klaers2010,Klaers2012}. Briefly, we use a dye-filled high-finesse microresonator made of two spherically curved mirrors at a $1.5\mu\textrm{m}$ distance to confine photons. The small mirror separation modifies the dye fluorescent emission such that the longitudinal mode number of the cavity photons is frozen out, and the system dynamics is restricted to the remaining two transverse mode numbers, making the photon gas effectively two-dimensional; see Fig.~\ref{fig1}(b). The transverse TEM$_{00}$ mode here is the energetically lowest allowed mode, and its frequency $\omega_\textrm{\tiny c}$ acts as a low-frequency cutoff, with $\hbar\omega_\textrm{\tiny c} \simeq 2.1\textrm{eV}$ in our case. The photon dispersion relation becomes quadratic as for a massive particle, and one finds that the mirror curvature leads to a harmonic trapping potential for the photon gas. By repeated absorption-emission cycles, a thermal population of cavity modes can be achieved when the dye fulfils the Kennard-Stepanov relation \cite{McCumber1964}, $B_{21}(\omega)/B_{12}(\omega)\propto \exp[-\hbar(\omega-\omega_\textrm{\tiny ZPL}/k_{\textrm{\tiny B}} T]$, where $\omega_\textrm{\tiny ZPL}$ denotes the frequency of the zero-phonon line of the dye; see also Fig.~\ref{fig1}(b). This Boltzmann-type scaling of the ratio between the Einstein coefficients for absorption and emission $B_{12}(\omega)$ and $B_{21}(\omega)$, respectively, can be understood from a thermalization of the rovibrational dye manifold due to frequent (fs timescale) collisions between dye and solvent molecules. In our experiment, the dye reservoir is pumped with an external laser beam, to provide an initial photon population. The pumping is maintained throughout the experimental cycle, as our system is not a perfect "photon box" due to losses from optical modes not confined in the cavity, nonradiative decay, and finite mirror reflectivity, which must be compensated for. In earlier work, we verified both the thermalization and Bose-Einstein condensation of the photon gas above a critical photon number \cite{Klaers2010th,Klaers2010}.

Because of the effective particle exchange between the photon gas and the dye reservoir by absorption and emission processes, we expect that grand-canonical conditions can be realized in this system for a sufficiently large dye reservoir \cite{Klaers2012}. For a finite size reservoir, however, deviations from a truly grand-canonical behavior will come into play. The particle exchange determines the photon number distribution within each cavity mode, both above and below the BEC phase transition. The expected probability $P_n$ to find $n$ photons in the resonator ground mode in thermodynamic equilibrium has been derived both by solving the master equation \cite{Klaers2012} and by employing a hierarchical maximum entropy calculation \cite{Sobyanin2012}. Assuming that the Kennard-Stepanov relation is fulfilled, one finds a photon number distribution
\begin{equation}
P_n=P_0 \frac{(M-X)!X!}{(M-X+n)!(X-n)!}\left(e^{-\hbar\Delta/k_\textrm{\tiny B}T}\right)^n,
\label{photondist}
\end{equation}
where $M$ is the number of dye molecules contributing to the reservoir, the excitation number $X$ is the sum of photons in the ground mode and the number of electronically excited molecules, and $\Delta=\omega_\textrm{\tiny c}-\omega_\textrm{\tiny ZPL}$ denotes the dye-cavity detuning. Equation~(\ref{photondist}) interpolates between two statistical regimes: for a large number of dye molecules and a small detuning, the distribution is Bose--Einstein-like (Ògrand-canonical caseÓ), with the most probable photon number being $n = 0$ and particle number fluctuations $\delta n_0$ of order of the average photon number in the condensate mode $n_0$. On the other hand, for decreasing reservoir sizes and larger detunings (or larger condensate fractions), the reservoir no longer supports large fluctuations and the distribution becomes Poisson-like (Òcanonical caseÓ) with a maximum at a nonvanishing photon number and a reduced mean square deviation \cite{Klaers2012,Sobyanin2012}. For example, relative fluctuations of $\delta n_0/n_0 = 75\%$ are expected for
\begin{equation}
n_0^2\simeq M_\textrm{\tiny eff} = M/\left[2+2\cosh(\hbar\Delta/k_\textrm{\tiny B} T)\right].
\label{crossover} 
\end{equation}
Grand-canonical statistics is thus expected to become important if an effective reservoir size $M_\textrm{\tiny eff}$ is quadratically larger than $n_0$. In general, large fluctuations are expected for high dye concentrations and a cutoff frequency near the dye zero-phonon line.

To detect intensity correlations of the condensed photons, we rely on a Hanbury Brown--Twiss set-up; see Fig.~\ref{fig1}(a). To separate the condensate mode from higher transverse modes, part of the radiation transmitted through one cavity mirror is spatially filtered in the far field (transverse momentum filter) and subsequently imaged onto an aperture located in an image plane of the microresonator (position filter), which efficiently suppresses the contributions of higher transverse modes \cite{note}.
The transmitted light passes through a polarizer to remove the twofold polarization degeneracy of the condensate mode and is then directed to the Hanbury Brown--Twiss setup. Here, the light is split into two paths and directed onto avalanche photodiodes (APDs) with single-photon sensitivity. By recording time histograms of detection events with an electronic correlator, time correlations of the condensate population can be determined with a temporal resolution of $\simeq 60\ \textrm{ps}$. For that, we evaluate the time-dependent second-order correlation function of the ground mode defined by $g^{(2)}(t_1,t_2)=\langle n_0(t_1)n_0(t_2)\rangle/\langle n_0(t_1)\rangle \langle n_0(t_2)\rangle$. A typical data set is shown in Fig.~\ref{fig1}(c). We find that the average condensate population is to good approximation time independent within one measurement cycle (pump pulse with duration $\simeq 150\ \textrm{ns}$), and the second-order correlations only depend on the time delay $\tau = t_2-t_1$. Accordingly, further analysis is performed with the time-averaged second-order correlation function $g^{(2)}(\tau)=\langle g^{(2)}(t_1,t_2) \rangle_{t_2-t_1=\tau}$.

\begin{figure}
\includegraphics{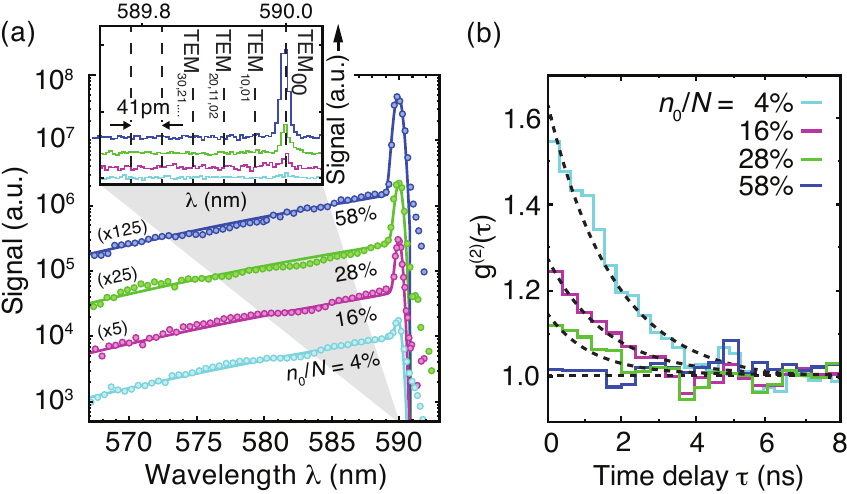}
\caption{(a) Spectral photon distribution for different condensate fractions (circles), each following a $300\ \textrm{K}$ Bose-Einstein distribution (solid lines). The inset shows corresponding high-resolution spectra and the position of the transverse cavity modes, revealing macroscopic occupation of the condensate mode only. In both graphs, data curves have been vertically shifted for clarity. (b) Corresponding second-order correlation functions $g^{(2)}(\tau)$ show photon bunching up to high condensate fractions. Experimental parameters: condensate wavelength $\lambda_\textrm{\tiny c}=2\pi c/\omega_\textrm{\tiny c}=590\textrm{nm}$ ($\hbar\Delta = -6.7k_\textrm{\tiny B}T$), dye concentration $\rho=10^{-3}\textrm{mol/l}$ (rhodamine 6G). \label{fig2}}
\end{figure}

Typical measurement results for a fixed size of the molecular reservoir, i.e., a fixed dye concentration and dye-cavity detuning, are presented in Fig.~\ref{fig2}. First, the thermodynamic state of the photon gas at room temperature is determined from the spectral distribution of light emitted from the cavity. The spectra shown in Fig.~\ref{fig2}(a) all are in the condensed phase with an average photon number beyond the critical particle number of $N_\textrm{\tiny c} \approx 90~000$, and exhibit a condensate peak at the wavelength of the cavity cutoff along with a thermal wing. We obtain the condensate fraction $n_0/N$ by a fit to a $300~\textrm{K}$ Bose-Einstein distribution (see the Supplemental Material\cite{supplementary}). Additionally, we verify the single-mode property of the condensate by directing part of the cavity emission through a high-resolution spectrometer; see the inset of Fig.~\ref{fig2}(a). The observed ground mode spectral width of $9~\textrm{pm}$ is clearly below the $41~\textrm{pm}$ spacing between transverse cavity modes TEM$_{\tiny mn}$ (dashed lines), which allows us to spectrally resolve the condensate mode. Results for the second-order correlation function $g^{(2)}(\tau)$ are shown in Fig.~\ref{fig2}(b). We observe photon bunching with $g^{(2)}(0) > 1$ followed by an exponential decay to $g^{(2)}(\tau) \simeq 1$ at large delays, while an immediate dropoff to $g^{(2)}(0) = 1$ at the BEC transition would be expected in the case of strictly conserved particle number \cite{Klaers2012,Naraschewski1999}. The bunching extends into the condensed phase up to a certain condensate size. 

\begin{figure}
\includegraphics{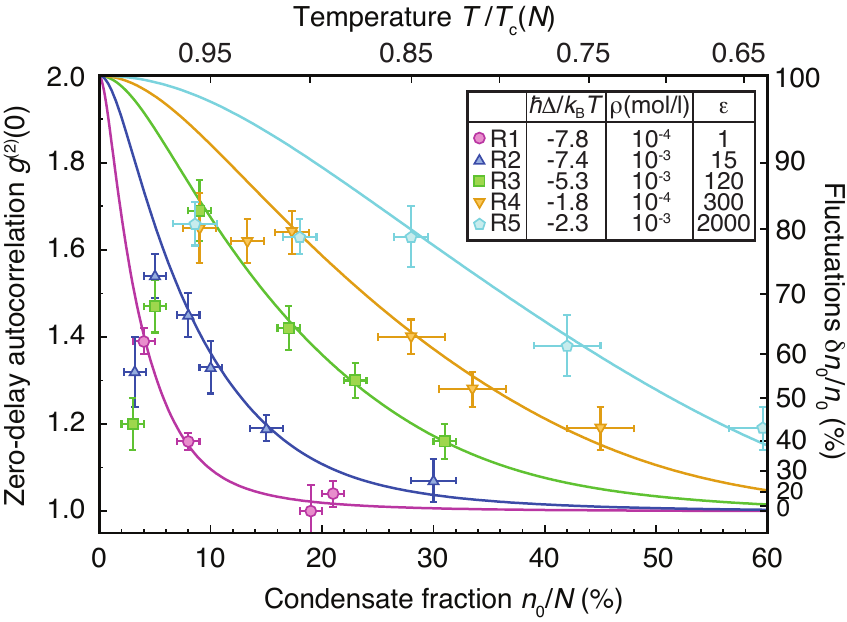}
\caption{Zero-delay autocorrelations $g^{(2)}(0)$ and condensate fluctuations $\delta n_0/n_0$ versus condensate fraction $n_0/N$ (or the corresponding reduced temperature $T/T_\textrm{\tiny c}(N)$ at $T=300\ \textrm{K}$), for five different reservoirs. An increase of the effective molecular reservoir size is quantified by an increasing value of $\varepsilon$, defined in Eq.~(\ref{reservoirsize}), with the corresponding values given in the third column of the table. Condensate fluctuations extend deep into the condensed phase for high dye concentration $\rho$ and small dye-cavity detuning $\Delta$ (R5). Results of a theoretical modeling are shown as solid lines. The error bars indicate statistical uncertainties. Experimental parameters: condensate wavelength $\lambda_\textrm{\tiny c}=2\pi c/\omega_\textrm{\tiny c} =\{598,595,580,598,602\}\textrm{nm}$ for data sets R1--R5; dye concentration $\rho=\{10^{-4},10^{-3},10^{-3}\}\textrm{mol/l}$ for R1--R3 (rhodamine 6G) and $\rho =\{10^{-4},10^{-3}\}\textrm{mol/l}$ in R4 and R5 (perylene red). For the theory curves, we find reservoirs sizes $M=\{5.5 \pm 2.2, 20.5 \pm 7.1, 16.0 \pm 5.7, 2.1 \pm 0.4, 10.8 \pm 3.7\}\times10^9$ for R1--R5, respectively. \label{fig3}}
\end{figure}

To verify the grand-canonical nature of the system in the condensed regime, we have varied the reservoir size. Figure 3 shows the zero-delay correlations $g^{(2)}(0)$ against the condensate fraction $n_0/N$ for five combinations of dye concentration $\rho$ and dye-cavity detuning $\Delta$. The data sets labeled with R1--R3 were obtained with rhodamine 6G dye (zero-phonon line $\omega_\textrm{\tiny ZPL} \simeq 2\pi c/545~\textrm{nm}$). For measurements R4 and R5, we have used perylene red ($\omega_\textrm{\tiny ZPL} \simeq 2\pi c/585~\textrm{nm}$) as a dye species that allows us to reduce the detuning between condensate and dye reservoir, effectively increasing the reservoir size. To quantify the relative effective reservoir size, we give the ratio
\begin{equation}
\varepsilon = \frac{M_{\textrm{\tiny eff},R_i}}{M_{\textrm{\tiny eff},R1}} = \frac{\rho_{R_i}}{\rho_{R1}}\times\frac{1+\cosh(\hbar\Delta_{R1}/k_\textrm{\tiny B} T)}{1+\cosh(\hbar\Delta_{R_i}/k_\textrm{\tiny B}T)}
\label{reservoirsize}
\end{equation}
in the third column of the table in Fig.~\ref{fig3}, where we normalize to the measurement with reservoir R1; see also Eq.~(\ref{crossover}). For the lowest dye concentration and largest detuning (R1), the particle reservoir is so small that condensate fluctuations are damped almost directly above the condensation threshold ($N \geq N_\textrm{\tiny c}$). By increasing dye concentration and decreasing the dye-cavity detuning one can systematically extend the regime of large fluctuations to higher condensate fractions. For the largest reservoir realized (R5), we observe zero-delay correlations of $g^{(2)}(0) \simeq 1.2$ at a condensate fraction of $n_0/N \simeq 0.6$. At this point, the condensate still performs large fluctuations of $\delta n_0/n_0 = (g^{(2)}(0)-1)^{1/2} \simeq 45\%$, although its occupation number is comparable to the total photon number of $N \approx 240~000$. This is seen as clear evidence for photon statistics determined by grand-canonical particle exchange between condensate and dye reservoir. Our results are well recovered by a theoretical modeling shown as solid lines in Fig.~\ref{fig3}, except for small condensate fractions below $\simeq 5\%$. This is attributed to imperfect mode filtering, which does not fully preclude photons in higher transverse cavity modes from reaching the correlation system with the corresponding averaging over modes causing a dropoff of the correlation signal at this point. In addition, when inspecting the data with largest number fluctuations, corresponding to reservoirs R4 and R5, we observe a limiting maximum value for the condensate bunching in the region $g^{(2)}(0) \simeq 1.6--1.7$, leading to a deviation from theory in those measurements. For each theory curve shown in Fig.~\ref{fig3} a single fitting parameter is used, the reservoir size $M$, which is determined by comparison with experimental results (Supplemental Material \cite{supplementary}). Good agreement of the theory curves with our experimental data is obtained if the reservoir size is assumed to be on the order of $10^9-10^{10}$ molecules. This suggests that not only dye molecules located in the TEM$_{00}$ mode volume of the resonator ($10^8$ molecules for a dye concentration of $10^{-3}~\textrm{mol/l}$), but rather a larger effective dye volume size contribute to the reservoir, as understood from a coupling between molecules in the ground mode volume and the whole molecular reservoir due to the emission and absorption of transversally excited photons (see the Supplemental Material \cite{supplementary}). Losses and pumping can in principle cause deviations from the expected statistical distribution, but we do not observe corresponding effects at our experimental conditions, where photon reabsorption dominates over photon loss \cite{Klaers2012,Keeling2013,Stoof2013}. The observed dependence of the photon statistics on reservoir size and detuning clearly demonstrates that saturation of the reservoir, as predicted from grand-canonical BEC theory, determines the observed photon statistics. 

\begin{figure}
\includegraphics{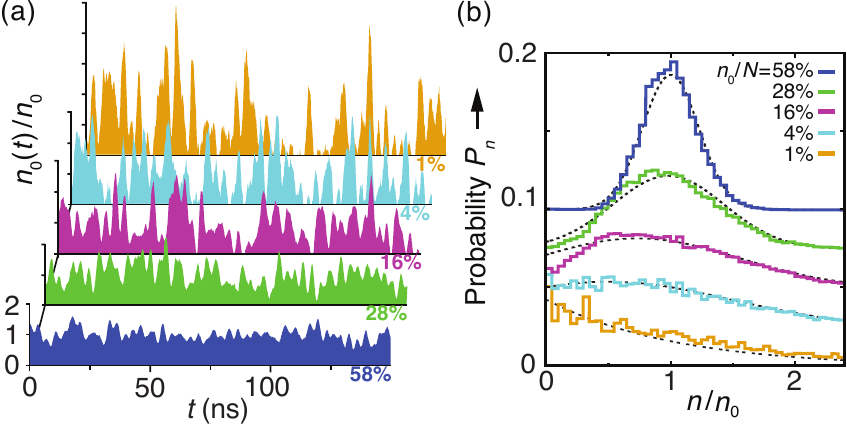}
\caption{(a) Temporal evolution of the normalized condensate mode population $n_0(t)/n_0$ ($\simeq 2\textrm{ns}$ temporal resolution). For increasing condensate fraction a damping of the fluctuations is observed. (b) Corresponding photon number distributions showing a crossover from Bose--Einstein-like to Poisson-like statistics. For illustration, the data curves have been vertically shifted and dashed lines present results from a theoretical model. Experimental parameters are as in Fig.~\ref{fig2}.\label{fig4}}
\end{figure}

The measured correlation time of approximately $\tau_c^{(2)} \simeq 2~\textrm{ns}$, see Fig.~\ref{fig2}(b), is sufficiently long to directly monitor the time evolution of the photon condensate with a photomultiplier tube, though such a measurement is affected by the finite ($2~\textrm{ns}$) instrumental photomultiplier bandwidth. Corresponding data for a fixed reservoir size are presented in Fig.~\ref{fig4}(a), showing a strongly fluctuating intensity for small condensate fractions, and weaker fluctuations when the condensate occupation increases. From this data, it is, moreover, possible to determine the photon statistics; see Fig.~\ref{fig4}(b). The obtained photon number distributions show a crossover from a nearly exponentially decaying (Bose--Einstein-like) distribution for the strongly fluctuating condensate to a Poisson-like distribution in the regime with small fluctuations. The measurements are in good agreement with the photon number distributions from the theoretical model (dashed lines), see Eq.~(\ref{photondist}), which describes the crossover from canonical to grand-canonical behavior.

To conclude, we have observed anomalously large particle number fluctuations in photon Bose-Einstein condensates over a wide range of condensate fractions, due to grand-canonical particle exchange of the condensate mode with a reservoir. At high quantum degeneracy, when almost all photons condense into the ground state of the system, this yields experimental evidence for the grand-canonical fluctuation catastrophe. The results, moreover, contribute to a clarification of the relation between Bose-Einstein condensates and the laser. As demonstrated by our measurements, the emergence of second-order coherence does not necessarily coincide with BEC criticality, and can be delayed to high condensate fractions. If second-order coherence is considered as an essential property of laser light, which is a common point of view in laser physics \cite{Siegman1986,Wiersig2009}, the number fluctuations in the Bose-Einstein condensate are a distinguishing property towards a laser. For other laser definitions, also laser light without second-order coherence can also occur \cite{Lien2001,Jakeman1970,Hofmann2000}. This ambiguity does not occur in a thermodynamic system, as in the here-investigated photon Bose-Einstein condensate, where a well-defined phase transition (the BEC transition) separates the thermal from the condensed phase.

For the future, it will be interesting to explore the effects of interactions on the condensate fluctuations and investigate superfluid properties of the system. Photon self-interactions \cite{Klaers2010}, which become increasingly important for large condensate sizes, can eventually provide a limit to condensate fluctuations.

\bibliography{grandcanonical.bib}

\end{document}